# Robust Excitonic Coherence Driven by In-Plane Anisotropy in a van der Waals Semiconductor

*Rup Kumar Chowdhury*, Md Samiul Islam, Marie Barthelemy, Nicolas Beyer, Lorry Engel, Jean-Sebastien Pelle, Mircea Rastei, Alberto Barsella, and Francois Fras* *

*Université de Strasbourg, CNRS, Institut de Physique et Chimie des Matériaux de Strasbourg, UMR 7504, F-67000 Strasbourg, France*

**Abstract**

The discovery of in-plane anisotropic excitons in van der Waals semiconductors enables state-of-the-art nanophotonic applications. A key parameter of these quasiparticles is the coherence time ($T_2$), which measures quantum dephasing and determines how long coherent superpositions can be preserved—critical for quantum photonic and excitonic technologies.

In particular, such systems pose fundamental questions about how anisotropy influences coherence properties. Here, we employ transient micro–four-wave mixing spectroscopy to study the coherence and population dynamics ($T_1$) of excitons at resonance in pristine layered rhenium disulfide. We probe and quantify the dephasing induced by many-body excitonic scattering processes, revealing signatures of anisotropy and one-dimensional features. This results in notable decoherence resilience against variations in optical excitation density and temperature, enabling measurables quantum features even at room temperature. Additionally, the absence of photon echo signals highlights the homogeneous nature of excitonic transitions in $ReS_2$ and reflects a low level of disorder across few-layer to bulk-like flakes. These findings open further exploration of in-plane anisotropy–driven decoherence in van der Waals materials, while creating valuable possibilities for quantum-based applications.

## Introduction

Layered van der Waals (vdW) semiconductors can be processed to an atomically thin level, with impressive structural and optical properties, including chemical stability and mechanical flexibility. Their exceptional exciton binding energies enable stable Coulomb-bound excitons even at room temperature [1–3]. Their two-dimensional (2D) nature also allows easy integration into photonic chips [4,5] and heterostructure devices [6,7]. At a fundamental level, the difference between strong in-plane covalent bonds and weak out-of-plane vdW interactions underpins the inherent out-of-plane anisotropy of vdW materials. In contrast, in-plane anisotropy, which is an alluring but relatively rare phenomenon, arises naturally in a few unconventional vdW systems [8–10]. Among them, black phosphorus (BP) and Re-based ($ReX_2$, X=S, Se) layered semiconductors has recently garnered significant research interest due to their strong in-plane anisotropic excitonic properties, enabling novel applications particularly in the domain of polarization-sensitive optoelectronics [11–14]. In $ReX_2$, owing to the reduced lattice symmetry, Jahn-Teller-like structural distortion, and zigzag chain-like structure, a pair of highly polarization-sensitive anisotropic excitons emerge naturally in their 2D planes [15–17] due to the valence band structure [18]. This anisotropic excitonic configuration gives rise to a rich array of emergent phenomena, including quasi-one-dimensional excitonic characteristics in 2D few-layer or bulk-like hosts [8,19–22] even without strain engineering [23], anisotropic moiré excitons [24], quantum beating [25], and the ability to host self-hybridized exciton-polaritons [26,27] with potentially an hyperbolic dispersion [28]. Furthermore, the low symmetry and strong anisotropy are theoretically well-suited to exhibit exciton superfluidity [29]. While exhibiting similar anisotropic attributes, Re-based systems offers an important advantage over BP in terms of environmental stability [30]. Unlike BP, the anisotropic excitonic properties of Re-based samples can be probed in their pristine form, i.e., in absence of encapsulation, making them beneficial for practical applications.

Unlike the more extensively studied group-VI transition metal dichalcogenides (TMDCs), namely $MX_2$ (M = Mo, W; and X = S, Se), which exhibit direct band exciton transitions only in their monolayer limit [1,2], anisotropic vdW materials such as Re-based TMDCs maintains strong excitonic response even in multilayer form [31], indicating a band structure that is weakly dependent on layer thickness. Yet, the fundamental question of whether the gap is truly direct or indirect is the subject of active debate to date [32,33]. In this context, time-resolved and nonlinear spectroscopy can provide valuable insights [34]. Earlier reports with time-resolved non-resonant approaches indicate exciton lifetime in ReS2 within 10-100 ps [35–39]. However these methods cannot infer sub-picosecond dynamics due to either limited time resolution or the involvement of charge relaxation pathways leading to exciton formation, potentially hindering the exploration of intrinsic dynamics at the excitonic levels.

Furthermore, and of particular importance, in anisotropic van der Waals materials, key aspects such as excitonic coherence properties (i.e., excitonic dephasing or homogeneous linewidth $\gamma$), the role of many-body scattering effects, and the extent of excitonic disorder (i.e., the inhomogeneous contribution to dephasing) remain largely unexplored. These factors govern the duration $T_2$ of the coherent superposition of excitonic states (related via $T_2 = 2\hbar/\gamma$), the ability to manipulate exciton states through coherent interactions [[35]], and directly influence the efficiency of energy transfer change transport within excitonic systems [[40,23,41]]. Understanding and controlling these factors is then crucial for optimizing excitonic device performance and advancing quantum technologies. More specifically in such anisotropic vdW materials, the interplay between strong in-plane anisotropy, the particularly large predicted exciton binding energy [[21,9]], and the emergence of one-dimensional excitons raises critical questions about their impact on coherence properties and the excitonic scattering processes.

To address these questions, resonant four-wave mixing (FWM) spectroscopy [[42–44]] is particularly suitable, providing a direct pathway to investigate the dynamics of exciton coherence and population by resonantly exciting specific optical transitions. However, the exciton properties of anisotropic vdW materials can depend not only on the flake thickness and the location within the flake, but also on the orientation of individual micrometer-scale flakes [[15,16]]. Therefore, a microscopic approach is essential. In this context, combining FWM time-resolved spectroscopy with microscopy presents a notable challenge.

Here, we performed state-of-the-art FWM micro-spectroscopy based on optical heterodyning and spectral interferometry [[45]] on layered $ReS_2$. This experimental configuration allows to probe areas with spatial resolution limited by diffraction and a temporal resolution determined by pulse duration (∼100 fs). We conducted a comprehensive investigation into the coherence and population dynamics of the two lower-level anisotropic excitons, examining the effects of key parameters such as: (i) temperature, (ii) optical power, (iii) flake locations and thickness from mono-layer to bulk-like. The analysis of pure excitonic dephasing enables us to probe and quantify exciton–phonon and exciton–exciton scattering, uncovering signatures of anisotropy and one-dimensional features driving a particular resilience against decoherence. Furthermore, examining both the dephasing and inhomogeneous broadening allows us to conclude that the excitonic linewidth is homogeneously limited, indicating a low level of excitonic disorder in bare $ReS_2$ — from few-layer to bulk-like flakes. These unique characteristics distinguish them from other isotropic 2D semiconductors. Conversely, the study of exciton population dynamics using our resonant approach reveals multiple exponential decay components, which comprise an ultrafast one (~150 fs), and involve scattering between bright and dark excitonic states.

## Results

$ReS_2$ sample was prepared *via* mechanical exfoliation (see Methods). Six different flakes were investigated, identified in white light imaging by their contrast (Fig. 1b), and in atomic force microscopy by their thickness ranging from bulk-like (Flake-1) to monolayer (Flake-6) (Fig. 1c and Fig. S1, supplementary information (SI)). The optical emission from layered $ReS_2$ is dominated by excitonic transitions [16,31]. The exciton resonances are first identified from the hyperspectral PL imaging (Fig. 1d) of which a spectrum example is presented in Fig. 1e, featuring two peaks associated with the anisotropic excitons $EX_{1(2)}$ at ~1.54 (1.57) eV.

Following the PL characterizations of $EX_{1,2}$, we performed FWM micro-spectroscopy (details in Methods, SI-Sec. 2, and [45]) to investigate their coherent dynamics (Fig. 2). FWM arises from third-order nonlinearity involving three interactions between the optical pulses and the system. The first interaction induces a coherent polarization in the sample, which is converted into a population by the second interaction, and the third one interacts with the latter population to yield the FWM signals. We employed two configurations. In the first (Fig. 2a), we utilized two pulses ($E_{1,2}$), leading to a FWM field $\propto \mu^4 E_1^* E_2^2$, with $\mu$ the electric dipole moment. $E_{1,2}$ are temporally separated by a delay ($\tau_{12}$), allowing to retrieve the dynamics of excitonic coherences. In the second configuration (Fig. 2b), three pulses ($E_{1,2,3}$) were employed with $E_1$ and $E_2$ overlapping ($\tau_{12} \approx 0$). The FWM signal $\propto \mu^4 E_1^* E_2 E_3$ is monitored as a function of the delay ($\tau_{23}$) between $E_2$ and $E_3$, enabling to retrieve the dynamics of excitonic population. The heterodyne detection technique allows FWM to be conducted with collinear pulse beams by radiofrequency ($\Omega_{1,2,3}$) shifting the pulse trains. The FWM field is subsequently detected via spectral interference with a reference beam, as exemplified in Fig. 2d.

***Coherence dynamics***. To optimize the FWM signal for each exciton $EX_{1,2}$, we first analyzed the linear polarization-resolved FWM signal in the two-pulse configuration with a fixed $\tau_{12}$, and determined the local dipole orientations under resonant conditions for excitons $EX_{1,2}$ (Fig. 2e). The signal exhibits a $cos^4(\theta - \theta_0)$ behaviour, where $\theta$ is the excitation polarization angle. This results from three interactions with electric fields, leading to FWM signal, and the fourth power from its projection onto the reference beam. We observe a twist of about 70° in the two linear dipoles, consistent with prior findings.

We now consider the excitonic coherent dynamics as a function of $\tau_{12}$. In this configuration, Four-wave mixing (FWM) spectroscopy offers the fundamental advantage of distinguishing between homogeneous broadening ($\gamma$), arising from irreversible dephasing (or coherence loss), and inhomogeneous broadening ($\sigma$), which results in apparent dephasing due to ensemble averaging over inhomogeneous excitonic energy distribution. We begin by evaluating the relative weight of $\sigma$ to $\gamma$ at 10 K, where it is expected to be the highest. To achieve this, the method is as follows: we initially examine the FWM dynamics in the 2D variables framework ($\tau_{12}$, $t$), obtained by Fourier transformation (see Methods), where a significant inhomogeneous contribution $\sigma$ should exhibit a clear photon echo signature. Figure 2f-g presents a 2D FWM plot at 10 K for both $EX_{1,2}$, for the multilayer Flake-2 (additional data for monolayer Flake-6 in Fig. S3, SI). Remarkably, we do not observe photon echo formation for both $EX_{1,2}$ as the maximum amplitude of the FWM signal does not follow the diagonal $t = \tau_{12}$. Instead, the signal evolves as free induction decay, suggesting a weak inhomogeneous contribution satisfying the condition $\sigma \ll \gamma$. Based on this initial estimation, the next step involves evaluating $\gamma$ and the upper bound of $\sigma$ by fitting the FWM amplitude as a function of $\tau_{12}$ (Fig. 2h-i). The fitting function accounts for the convolution of the pulse shape with the system response, characterized by an exponential decay of coherence ($\gamma$) and Gaussian-like inhomogeneous dephasing ($\sigma$). Further details of the fitting function are provided in SI-Sec. 3. For thick multilayer and monolayer $ReS_2$, $\gamma$ is around 5 meV and 8 meV respectively (Fig. 2h-i and Fig. S3, SI), while $\sigma$ remains negligible independently of flake thickness, with an upper bound of $\sigma \leq 1.0$ meV, as determined by the consistent fit quality up to this limit. The weak dependence of $\gamma$ on flake thickness at 10 K is shown in more detail in Fig. 3h. Complementary data at 77 K exhibits a similar trend (Fig. S5). We conclude that, even when thinning down to the monolayer limit, the exciton coherent dynamics in pristine $ReS_2$ are governed by the homogeneous contribution $\gamma$. In contrast, conventional isotropic semiconductor shows such behavior only when encapsulated in hBN [46–48]. Comparatively, in bare monolayer $MoSe_2$, exciton dynamics show a clear photon echo with a large $\sigma$ ranging between 7 to 15 meV [43].

We now present the detailed evolution of $\gamma$ (or coherence time, $T_2 = 2\hbar/\gamma$) as a function of (i) temperature, (ii) optical power, and (iii) flake locations, as summarized in Fig. 3 and Fig. S4. First, we explore the robustness of the excitonic coherences against sample temperature ($T$) and the dephasing induced by exciton-phonon interaction. Examples of FWM dynamics on multilayer Flake-2 with respect to temperature variations are shown in Fig. 3a-b, while $\gamma(T)$ for $EX_{1-2}$ is presented in Fig. 3c-d. Interestingly, $\gamma(T)$ exhibits notable quasi linear variations even up to room temperature, allowing extraction of the homogeneous limit $\gamma_0$ at zero temperature, approximately 3.9 ± 0.8 (5.8 ± 0.9) meV for $EX_{2(1)}$. Importantly, the coherent signature dynamics remain discernible, although approaching our time resolution limit, even at room temperature. This contrasts with $MX_2$ systems, where $\gamma$ becomes unresolved above

~120 K [43,46–49] (under similar experimental temporal resolution), highlighting $ReS_2$ specific resilience against temperature-induced decoherence. This pronounced linear behavior also differs from the well-reported parabolic trend of $\gamma(T)$ in $MX_2$ [43,46,47,49]. The thermal broadening of $\gamma$ is analyzed in more detail using a simple approach where $\gamma(T) = \gamma_0 + \alpha T + \beta\left(\exp\left(\frac{E_0}{k_B T}\right) - 1\right)^{-1}$, where 2nd and 3rd components represent the contributions of low-energy acoustic phonons and high-energy optical phonons at energy $E_0$, respectively. This arises from the phonon occupation number, with the linear dependence on $T$ for acoustic phonons due to the assumption that the energies of interacting acoustic phonons, constrained by momentum conservation, are less than the thermal energy $k_B T$. For both $EX_{1,2}$, the acoustic-phonon scattering value $\alpha$ is extracted around 33 ± 15 μeV/K. However, using this approach for the optical part, several pairs of $E_0$ and $\beta$ fit our data, with $E_0$ varying from 50 to 200 meV and $\beta$ from a few meV to 2.0 eV. Therefore, based on a recent estimation of $E_0$ [50], we constrained $E_0$ to around 50 meV and, from fitting, found it to be 61 (44) meV, with $\beta$ being 5.0 ± 2.6 (6.3 ± 1.1) meV for $EX_{1(2)}$, respectively. Finally, the extracted value of optical phonon coupling $\beta$ in $ReS_2$ is approximately 30 times lower than in $MoSe_2$ and $WSe_2$ monolayers [43,46] for similar optical phonon energy and acoustic phonon coupling, quantifying the particular robustness of $ReS_2$ excitonic coherence with exciton-phonon scattering induced decoherence. Such pronounced linear behavior of $\gamma(T)$, associated with relatively weak optical phonon coupling at resonance, mirrors findings in single carbon nanotube systems [51,52]. This behavior in layered $ReS_2$ may stem from the quasi one-dimensional aspect of excitons resulting from strong structural anisotropy [8,19–22]. The reduced dimensionality constrains available phase-space channels, potentially limiting exciton-phonon scattering. This aligns with experimental observations in 1D strain-engineered $WS_2$, where exciton propagation is enhanced while exciton-phonon scattering is significantly reduced [23]. A similar trend has been observed in ref. [53] where a direct comparison between $WSe_2$ and $WS_2$/$ReS_2$ heterostructures reveals enhanced valley polarization and a reduction in exciton-optical phonon scattering due to the presence of $ReS_2$ monolayers. However, further investigation is needed to refine this analysis, particularly regarding the limited role of optical phonons in the decoherence process as the exciton dimensionality is reduced [54].

Another key factor influencing the excitonic decoherence, is the exciton density, which governs the many-body exciton-exciton scattering rate. The exciton density can be directly tuned by adjusting the pump (E1) power density, $P$. Therefore, in Fig. 3e-f, we present the evolution of $\gamma$ as a function of $P$, where an increase in the homogeneous linewidth is observed with increasing $P$. This trend is commonly referred to as excitation-induced dephasing (EID) [55,56]. However, in the case of $ReS_2$, $\gamma(P)$ exhibits two distinct quasi-linear behaviors across different power regions. We then analysed the behaviour, in each region, using: $\gamma(P) = \gamma_{P0} + A \times P$, where $\gamma_{P0}$ is the density-independent dephasing rate (in the low power region I), and $A$ is the rate of EID. Under low power conditions (Region-I), $A$ is extracted around $5\times10^{-4}$ meVcm$^2$/W at 10 K. Complementary data obtained at 77 K shows similar trend (details of fits values provided in Fig. S6). It is challenging to directly compare the EID rate with those of other materials. However, if we base the comparison on energy per optical power density, the EID rate in region I, extracted, is an order of magnitude smaller compared to isotropic 2D materials $MX_2$ [47] (under similar pulses length and repetition rate). Importantly, the EID rate changes by more than two orders of magnitude between power regions I and II. Such a dramatic shift in the EID regime has not been observed in other 2D materials, such as isotropic TMDs or lead halide perovskites. This change could be consistent with a transition from a 1D- to a 2D-like exciton scattering type: at low power, anisotropy dictates the 1D behavior, while beyond a critical exciton density, 2D exciton-exciton interactions become dominant, as demonstrated in thin black phosphorus through exciton-exciton annihilation [57].

We conclude the study of excitonic coherence by exploring the influence of flake spatial location. Figure 3g shows $\gamma$ for both $EX_{1,2}$ across multiple locations on Flake-2 (spanning $\sim$10 μm), with weak spatial dispersion, indicating uniform excitonic behavior. This complements the low microscopic disorder (negligible $\sigma$) probed under diffraction-limited beams as discussed above.

***Population dynamics***. We next employ three-pulse ($E_{1,2,3}$) resonant excitation scheme (Fig. 2b) to probe the $EX_{1,2}$ population dynamics. At $\tau_{12} \approx 0$, the polarization selective resonant pumping creates an initial density of excitons within the light cone (Fig. 4a). These excitons subsequently relax through radiative and non-radiative channels. To analyse the interplay among these channels, we also recorded population dynamics as a function of power, temperature, and flake thickness (Fig S8-10). Figure 4b shows examples of population dynamics of $EX_1$ in with varying density and temperature. The dynamics suggest that population decays ($T_1$) $EX_{1,2}$ across multiple timescales, spanning from hundreds of femtoseconds to nanoseconds. Unlike in $MX_2$ [46,58], we did not observe any variation in the FWM phase with $\tau_{23}$, which would lead to population modulations. Thus, population dynamics are well-fitted using a multi-exponential transient response function, $R(\tau_{23}) \propto$ $\sum_{n=1}^{3} A_n \theta(\tau_{23}) exp\left(-\frac{\tau_{23}}{T_{1n}}\right)$, convoluted with a Gaussian pulse (SI Sec. 5), where $\theta(\tau_{23})$ represents the Heaviside function. This approach allows us to extract three distinct lifetime components ($T_{1n}$) and relative amplitudes $A_n/A$ for both $EX_{1,2}$ (Fig. 4c-d) under different experimental conditions (complementary data in SI, Sec. 6-7). To avoid ambiguities in the short-time-scale analysis, we conducted alternative curve fits (details in SI-Sec. 5) using i) only two measurable decay times, and ii) three decay channels, with the first considered instantaneous relative to the resolution ($T_{11}$ << pulse-duration), for instance due to a non-resonant contribution [36,38]. These fits resulted in lower quality, demonstrating the relevance of the three-resolvable-decay-channel analysis. For both excitons, we found an extremely fast initial decay channel $T_{11}$ with a characteristic time of approximately 150 fs, varying slightly with temperature, power, and flake thickness parameters. Following this, the second exponential component $T_{12}$ ranges from 0.5 to 1.0 ps and shows a similar relative amplitude to the initial component ($A_1/A \approx A_2/A$). The third decay timescale ($T_{13}$) exceeds nanosecond window for both excitons, with relative amplitude of $A_3/A \approx 2A_1/A$ at low power and temperature. Such ultrafast dynamics, featuring a 150 fs component, are consistent with previous findings on $MX_2$ systems performed at resonance [46,49,59], however it contrasts with non-resonant studies on $ReS_2$, which only highlighted dynamics in the range of 10-100 ps [36,38].

Importantly, the population dynamics ($T_{1n}$ and $A_n/A$) remain largely independent of flake thickness (see SI-Sec. 6-7), consistent with an unchanged band structure across varying numbers of layer [31]. To explore the mechanisms controlling population dynamics, we analyzed all three $T_1$ components as a function of optical power and temperature (Fig. 4b-d). With increasing optical power and temperature, all time components decrease. Additionally, the weight of the third channel decreases, while the proportion of the second channel ($A_2/A$) increases, becoming dominant at room temperature. Such multiscale dynamics suggest a strong connection between bright and dark state reservoirs [33,43,50,58,59] comprising excitons with in-plane momentum outside the light cone, and spin-forbidden dark states lying tens of meV below the bright states as recently revealed in $ReS_2$ [33]. Indeed, exciton scattering is highly efficient in TMDCs, particularly due to their relatively large exciton mass, wide exciton linewidth, and small conduction band splitting [1,2]. Therefore, bright excitons can significantly scatter into dark state reservoirs through non-radiative processes, such as exciton-phonon and exciton-exciton scatterings, including potentially exciton-exciton annihilation [57] and Auger processes [59]. With temperature and excitation density, these processes enhance their efficiency, effectively dominating the population dynamics. The different decay times of the dynamics result from a combination of the scattering processes between excited states and irreversible decay processes into the ground state. In this framework, the shortest component ($T_{11}$) arises from all processes that deplete the bright state, including radiative recombination and exciton scatterings from bright to dark states. The second ($T_{12}$) and third components ($T_{13}$) result from the population decay of the dark reservoirs, including the return of the exciton into the light cone via secondary scattering. Additionally, the population dynamics of $EX_1$ and $EX_2$ are notably similar across different temperatures and excitation densities (see Fig. 4c-d and SI-Sec. 6-7). This similarity, especially at low temperatures despite the 30 meV excitons splitting, suggests that inter-valley $EX_1$-$EX_2$ scattering plays a negligible role in population dynamics compared to other scattering mechanisms discussed.

## Conclusion

Concerning the nature of the band gap, comparison to established results for isotropic 2D semiconductors indicates that our findings are more consistent with $ReS_2$ exhibiting a direct band gap regardless of layer thickness[59]. Indeed, although the trend of coherence dephasing behavior differs significantly from that of isotropic semiconductors, our study reveals that $T_1$ values and $\gamma$ value at the lowest temperature and optical power, from monolayer to multilayers $ReS_2$, are comparable to those observed in isotropic $MX_2$ monolayers [42,43,46,47,49,58–60], which exhibit direct band gap characteristics. In contrast, indirect gap bilayer $MoSe_2$ [61] exhibits ultrashort $T_1$ components ( ~50 fs and 800 fs) and a broad $\gamma$ of approximately 30 meV at low temperature due to its complex indirect band structure that enables multiple ultrafast excitonic scattering pathways. Consistently, enhanced exciton–phonon scattering in $WS_2$ bilayers and multilayers has been reported [62], leading to an additional broadening of 25–30 meV at low temperature —features incompatible with our findings for $ReS_2$. Therefore, our results align more closely with the hypothesis that $ReS_2$ maintains a direct band gap regardless of layer thickness, which was suggested earlier using angle-resolved (*k*-space) photoemission spectroscopies on bulk [19] and mono- to bilayer layers $ReS_2$ [63].

In summary, we explored the intrinsic coherence and population dynamics of excitons in a pristine anisotropic vdW system ($ReS_2$), using ultrafast resonant four-wave mixing microscopy, across a range of experimental conditions. We investigated excitonic many-body interactions by probing the dephasing rate $\gamma$ as a function of excitation density and temperature, and found that anisotropy plays a key role in driving 1D excitonic features. This results in a particular decoherence resilience, compared to their isotropic counterparts, against exciton–exciton and exciton–phonon interactions, even allowing time-resolved excitonic coherence signatures at room temperature using ~100 fs optical pulses. This ophin excitonic systems ens up further in-depth exploration, particularly in understanding the relationship between excitonic anisotropy and decoherence while also creating new possibilities for quantum-based applications. Additionally, we found that the coherence and population exciton dynamics exhibit only a weak dependence on layer thickness. Furthermore, the inhomogeneous contribution is negligible (i.e., homogeneously dominated linewidth) from bulk-like to few-layer flakes, and the dephasing rate $\gamma$ remains nearly constant across large spatial regions within a flake. This points to a particularly low level of excitonic disorder in such anisoropic systems. Altogether, these features are key assets, for instance, in enabling efficient excitonic transport, regardless of flake thickness and free from the constraints of encapsulation processes.

## METHODS

***Sample fabrication and characterization.*** A commercially available quasi-transparent flat sapphire was used as the substrate, which was cleaned using Ar plasma (2 min) before proceeding with $ReS_2$ sample preparation. The $ReS_2$ flakes were mechanically exfoliated from a crystal using standard scotch-tape and directly transferred on the sapphire after thinning down. Following this, the identification of the flakes and characterization of their thicknesses were performed using a white light optical microscope and an atomic force microscope.

***PL and FWM micro-spectroscopy set-up.*** We developed a dual-mode experimental setup for PL and FWM spectroscopy, with confocal spatial mapping, using a 0.9 NA objective to focus the excitation spot to the diffraction limit. For PL measurements, we used a continuous-wave 532 nm laser to non-resonantly excite our layered $ReS_2$ sample. The excitonic emission was collected via a spectrometer (monochromator + CCD camera) with a spectral resolution of ~0.02 nm. The sample was placed inside a cryostat equipped with position-controlled piezoelectric nanopositioner stages. The optical pulses for FWM measurements were generated with a spectrally tunable Ti: sapphire laser oscillator with 80 MHz repetition rate. All pulses were then passed through a single-prism pulse compressor to minimize the pulse duration at the sample location, monitored with an autocorrelator (see SI Sec. 3). The temporal acquisition is achieved using a mechanical delay line that provides time-delayed collinear optical pulses. Linear polarizations, along with the spectral resolution, were used to address $EX_1$ and $EX_2$ individually. We separately measured the transient excitonic coherence and population signals using two-pulse ($E_{1,2}$) and three-pulse ($E_{1,2,3}$) sequence configurations, respectively. The FWM is detected using optical heterodyne and spectral interferometry [45,64]. The time-delayed excitation pulses, $E_{1,2}$ and $E_{1,2,3}$, were frequency upshifted by distinct radio-frequencies (RF) $\Omega_i$ around 110 MHz using acousto-optic modulators (AOM) drive by a custom-built RF mixer. The relatively weak FWM signal appearing at the FWM frequencies $2\Omega_2 - \Omega_1$ ($E_{1,2}$) or $\Omega_3 + \Omega_2 - \Omega_1$ ($E_{1,2,3}$), also around 110 MHz, differed from the excitation pulses radio frequencies $\Omega_i$. Then the signal was retrieved through spectral interference with a reference pulse, and detected using the spectrometer. The spectrally detected FWM signals, dependent on the pulse delay $\tau_{12}$ (coherence dynamics) or $\tau_{23}$ (population dynamics), were processed using fast Fourier transformations (FFT). This led to the extraction of both FWM phase and amplitude in either the time domain (single FFT) or frequency domain (FFT followed by a reverse FFT) [64]. The FWM amplitude is first explored in the 2D frame $(t, \tau)$, then integrated in $t$ and analyzed as a function of $\tau$ for each exciton, referred to in the main text as FWM dynamics (details are provided in SI section3).

## FIGURES

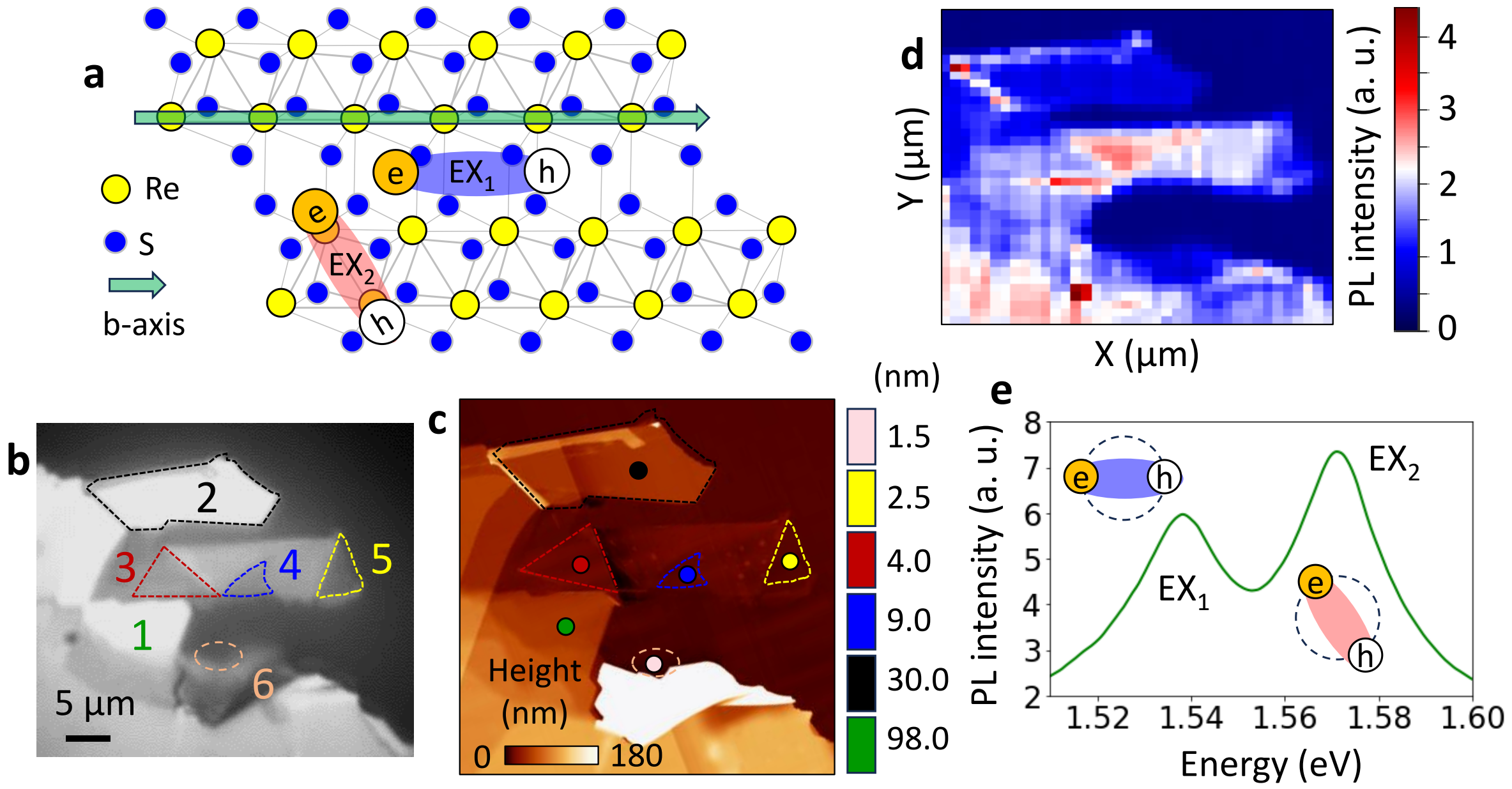

**Fig. 1 | Basic morphological and optical characterizations of layered $ReS_2$. a**, Schematic top view of a $ReS_2$ layer with in-plane anisotropic exciton pair ($EX_1$ and $EX_2$). Green arrow denotes the crystalline b-axis (along Re-Re chains). Yellow and blue dots characterize Re and S atoms, respectively. **b**, White light image of mechanically exfoliated $ReS_2$ flakes with different thickness. The different flakes studied are marked by the dashed lines and indicated by their flake numbers starting from Flake-1 to Flake-6. The scale bar is 5.0 μm. **c**, Atomic force micrograph of the $ReS_2$ flake locations with corresponding colour bar in the insets denoting height profiles of flakes marked with individual colours (vertical) and overall height profiles (horizontal). The thickness data extracted from AFM measurement for Flake-1 to 6 appeared as bulk-like (green), multilayers (black), few-layers (blue), 3-4 layered (brown), 2-3 layered (yellow), and 1-2 layered (pink) flakes, respectively. **d**, Hyperspectral PL map (77 K) (integrated over the entire spectral window range) of the flakes with the associated PL intensity colour scale. The step size is 1.0 μm. **e**, Typical PL spectrum taken on Flake-2, highlighting the two lowest energy optically bright excitons ($EX_1$ and $EX_2$) with schematic representation as shown in the inset.

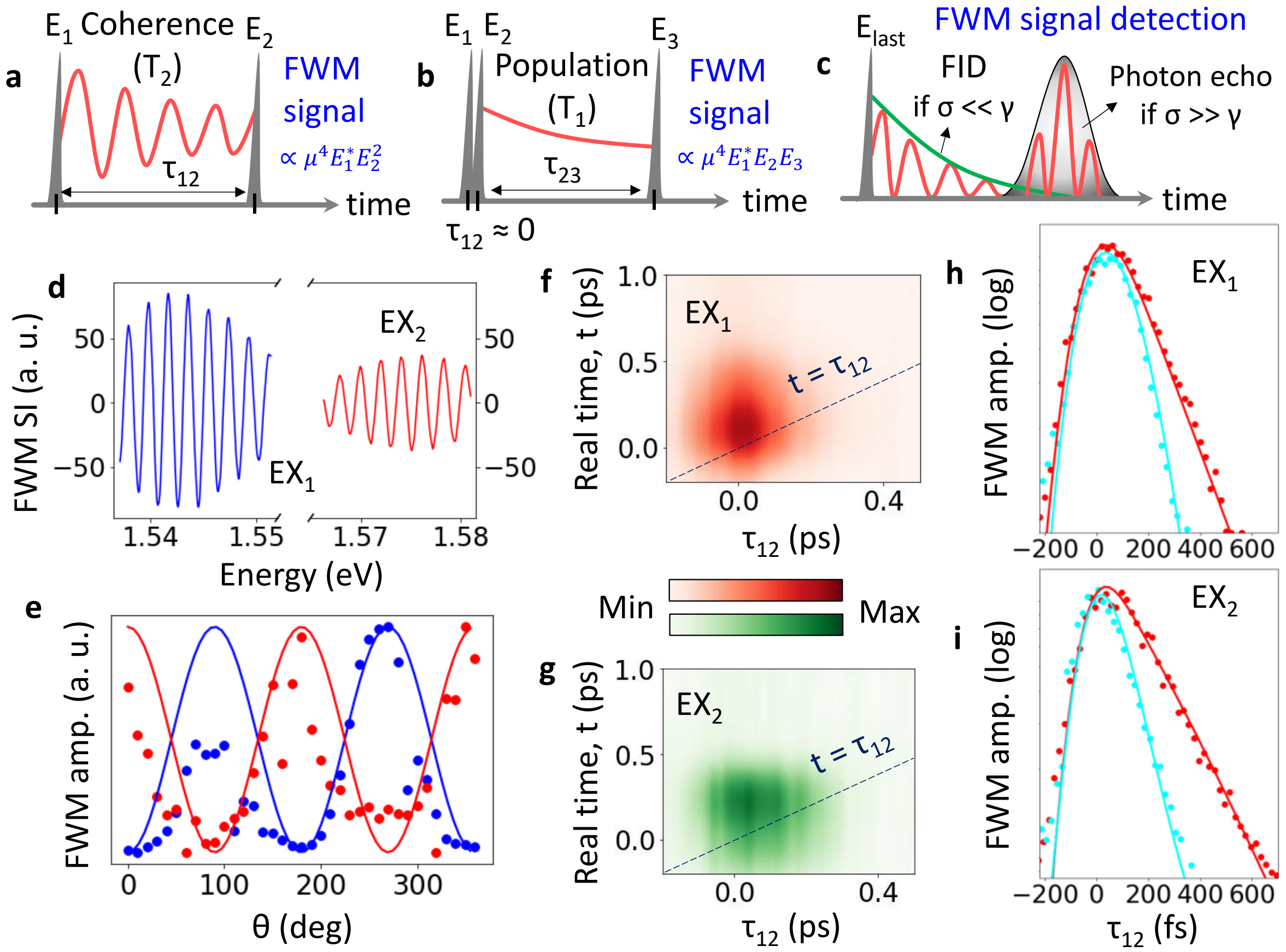

**Fig. 2 | Resonant FWM micro-spectroscopy on layered $ReS_2$.** **a**, Schematic representation of two pulse ($E_1$ and $E_2$) configurations, varying $\tau_{12}$ allow to probe the dynamics of coherences, characterised by coherence time $T_2$. **b**, Three pulse ($E_1$, $E_2$, and $E_3$) configurations, with $E_1$ and $E_2$ overlapping ($\tau_{12} \approx 0$), varying $\tau_{23}$ allows to probe the dynamics of exciton population. The description of the pulse sequence on the system is provided in the main text and method part. **c**, Representation in real time, $t$ of FWM for two situations: when homogenous contribution dominates called free induction decay (FID) and when inhomogeneous contribution dominates called photon echo, where the maximum of the echo is at $t = \tau_{12}$. $E_{last}$ in the schematic diagram represents $E_2$ in two pulse and $E_3$ in three pulse configurations. **d**, Example of FWM signal: spectral interferograms of $EX_1$ (blue line) and $EX_2$ (red line) obtained at resonant excitonic energies recorded for a specific delay value $\tau_{12}$. From these interferograms, the FWM amplitude, either in time or spectrum, is extracted through Fourier analysis (see Methods and SI) for each delay $\tau_{12}$. **e**, FWM amplitude, at $\tau_{12} \approx 0$, as a function of $E_{1,2}$ linear polarization angle (identical for both beams) for both $EX_1$ (blue dots) and $EX_2$ (red dots) with the corresponding fits (solid lines). FWM amplitudes resolved in function of real time $t$ and delays $\tau_{12}$ for **f**, $EX_1$ and **g**, $EX_2$ with corresponding colour scale for multilayer Flake-2. Examples of

FWM amplitudes versus $\tau_{12}$ presented for **h**, $EX_1$ and **i**, $EX_2$ at 10 K and 50 W/cm$^2$ along with the corresponding fits (solid lines) for monolayer (cyan) and multilayer (red) $ReS_2$ flake.

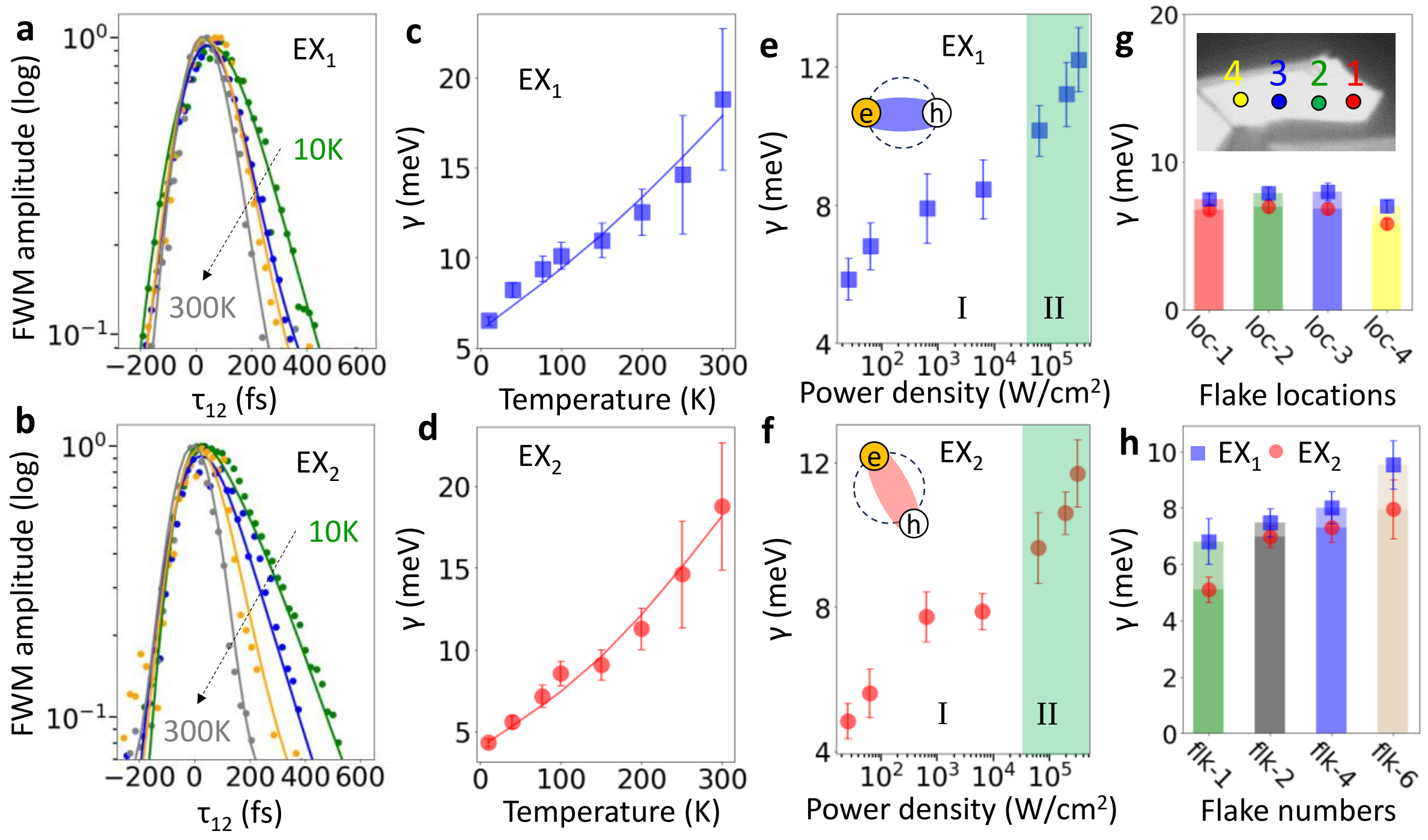

**Fig. 3 | Coherent dynamics of anisotropic excitons in layered $ReS_2$.** Examples of FWM amplitudes versus $\tau_{12}$ presented for **a**, $EX_1$ and **b**, $EX_2$ at several temperatures starting from 10 K to 300 K, along with the corresponding fits (solid lines) on Flake-2. Variation of $\gamma$, with corresponding fits, for **c**, **e**, $EX_1$ and **d**, **f**, $EX_2$ in function of temperatures and power density, respectively. Temperature dependent measurements shown in figure (**c**) and (**d**) performed at fixed pump power density 50 W/cm$^2$. Solid lines denote fits, with extracted parameters of temperature induced dephasing, are discussed in the main text. Conversely, power dependent measurements shown in figure (**e**) and (**f**) were conducted at a fixed temperature of 10 K. Two regions are marked based on power density: I with power density $< 10^4$ W/cm$^2$ (white shaded area) and II with power density $> 10^4$ W/cm$^2$ (green shaded area). Both regions exhibit distinct linear behaviours with different slopes. Inset of (**e**) and (**f**): schematic images of in-plane anisotropic excitons in $ReS_2$ with different dipolar orientations. **g**, Distribution of $\gamma$ for different locations on Flake-2. The inset shows different measurement locations on Flake-2, corresponding to colour codes of the bars. **h**, Homogeneous linewidth ($\gamma$) for $EX_1$ (blue) and $EX_2$ (red) as a function of flake numbers, where flake thickness decreases with increasing flake numbers as discussed in AFM analysis. Measurements in (**g**) and (**h**) were conducted at 10 K and 50 W/cm².

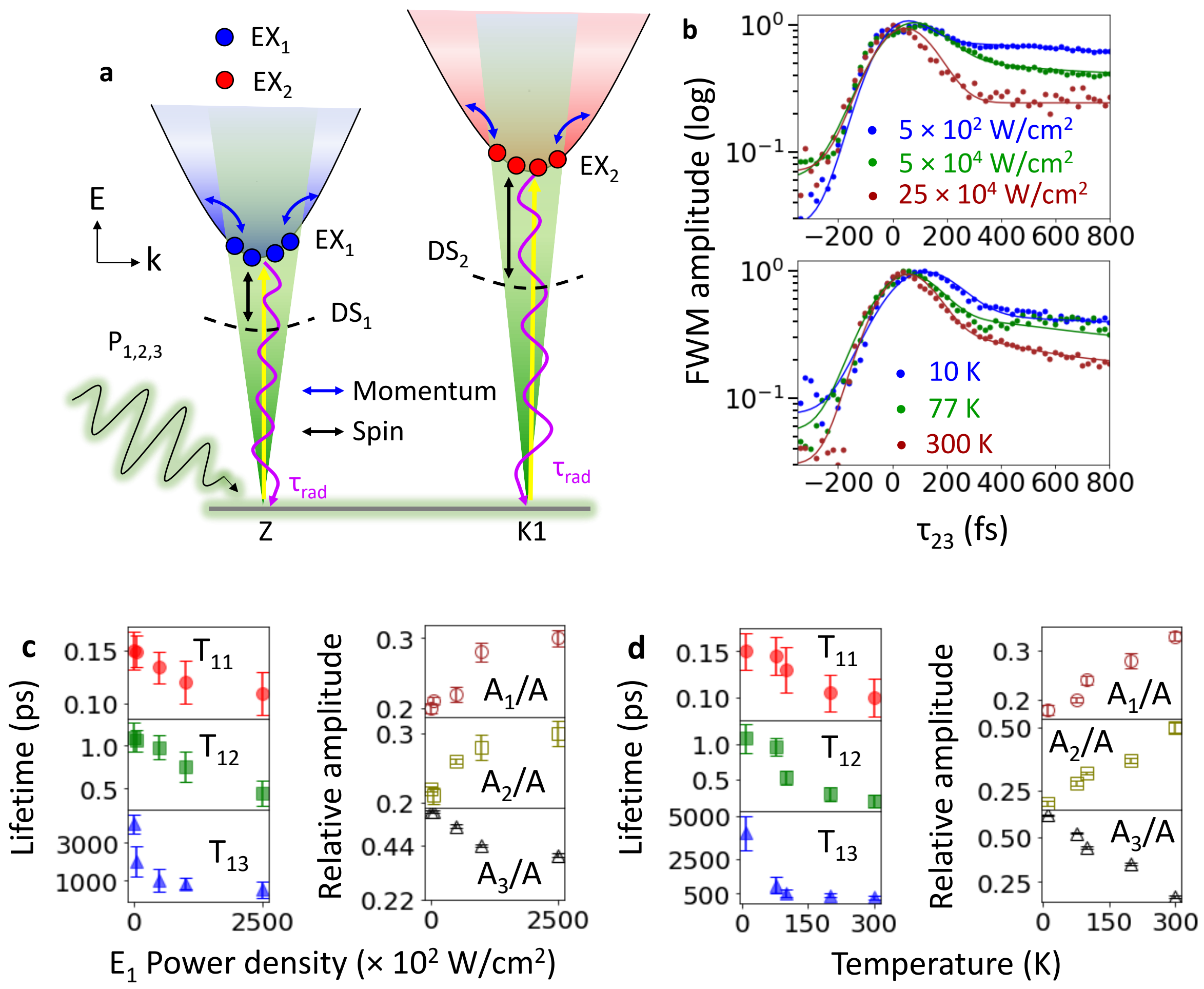

**Fig. 4 | Population dynamics of anisotropic excitons in layered $ReS_2$. a**, Schematic of the energy level dispersion (E, k), including optical transition and scattering processes, within the hypothetical framework of the two valleys (Z and K1) in $ReS_2$. The spin-allowed excitonic transition bands comprising the excitonic bright states $EX_{1,2}$ (blue and red solid circles), which lie within the light cones (green shaded areas) and are split by 30 meV (see Fig. **1e**). $DS_{1,2}$ designates the spin-dark states (no direct optical transition allowed), which, according to Ref. [[33]], are located 13 meV (20 meV) below $EX_{1(2)}$. The resonant optical excitations of $EX_{1,2}$ are depicted as yellow solid lines. From the light cone, $EX_{1,2}$ excitons can radiatively recombine (purple curved arrow), scatter to momentum-dark states outside the light cone (blue solid line), or scatter to spin-dark states $DS_{1,2}$ (black solid line). **b**, Examples of pump ($E_1$) power density and temperatures dependent population dynamics of $EX_1$ in terms of FWM field amplitude data with corresponding fits (solid lines) as function of the $E_{2,3}$ delay $\tau_{23}$. The power (temperature) dependent measurements are carried out at fixed temperature (power density): 10 K (50 W/cm$^2$). **c**, **d**, Summary of each transient FWM population dynamics fitting lifetimes (solid markers) and relative amplitudes (open markers) and corresponding errors represented

as error bars based on different parameters: power density and temperature for $EX_1$. Three different population timescales: $T_{11}$ (red), $T_{12}$ (green), and $T_{13}$ (blue) along with corresponding relative amplitudes $A_1/A$ (brown), $A_2/A$ (olive), and $A_3/A$ (black) of the decay processes are mentioned with different coloured markers. Here $A = A_1 + A_2 + A_3$. Note that the longest decay component (∼ns range) is accurately determined from the experiments conducted over an extended delay range, as shown in Fig. SI-S11. The data here were recorded on multilayer Flake-2.

## AUTHORS CONTRIBUTIONS

The project was originally proposed by FF. FF and MI built the experimental setup, with help from RC and MB. RC and MI carried out measurements, with help from MB and FF. RC and FF analysed the data with inputs from MI MB and AB. FF, RC, and MI conceived the analysis tools. RC fabricated the samples. NB, JP and LE provided a technical support. MR carried out the AFM measurements. FF and RC wrote the manuscript, with inputs from of MB MR and AB. FF supervised the project.

*Corresponding authors: fras@ipcms.unistra.fr, chowdhury@ipcms.unistra.fr

## FUNDING SOURCES

This research was supported by the French Agence Nationale de Recherche (ANR) under the grants FINDING ANR-18-CE30-0012-01 and LHNANOMAT ANR-19-CE09-0006, and from the QUSTEC program (European Union’s Horizon 2020, Marie Skłodowska-Curie Grant Agreement No. 847471).

## DISCLOSURES

The authors declare no conflicts of interest.